\documentclass{article}  

\usepackage{inputenc}
\usepackage{wrapfig}
\usepackage{hyperref}
\usepackage{graphicx}
\usepackage[demo]{rotating}
\usepackage{subfigure}
\usepackage[table,xcdraw]{xcolor}
\usepackage{adjustbox}
\usepackage{braket}
\usepackage{sectsty}
\usepackage{multicol}

\usepackage{float}
\usepackage{wrapfig}

\usepackage[margin=3.0cm]{geometry}
\usepackage{fancyhdr}
\usepackage[dvips]{epsfig}

\usepackage[dvips]{graphicx}

\usepackage{babel}

\usepackage[compact]{titlesec} 
\usepackage{listings}
\usepackage{amsmath}
\usepackage{siunitx}
\usepackage{notoccite}
\usepackage{setspace}

\title{Topological Superconductivity}
\author{Review Paper \\ Christian Stefan Gruber}
\date{February 2022}
\usepackage{abstract}
\usepackage[center]{caption}
\begin{document}





\begin{titlepage}

\begin{center}
 {\Large\textbf{Exploring Superconductivity: The Interplay of Electronic Orders in Topological Quantum Materials}\\ \vspace{0.5cm}}
 {\textbf{Christian Stefan Gruber$^{1}$, 
 Mahmoud Abdel-Hafiez$^{2,3,1}$}\\ \vspace{0.2cm}}
 \text{$^{1}$ Uppsala University, Department of Physics and Astronomy, Box 516, SE-751 20 Uppsala, Sweden}
 \text{$^{2}$University of Sharjah, Department of Applied Physics and Astronomy, P. O. Box 27272, UAE}
 \text{$^{3}$Lyman Laboratory of Physics, Harvard University, Cambridge, Massachusetts 02138, USA}
 \end{center}


\begin{abstract}

Topological quantum materials hold great promise for future technological applications. Their unique electronic properties, such as protected surface states and exotic quasiparticles, offer opportunities for designing novel electronic devices, spintronics, and quantum information processing. The origin of the interplay between various electronic orders in topological quantum materials, such as superconductivity and magnetism, remains unclear, particularly whether these electronic orders cooperate, compete, or simply coexist. Since the 2000s, the combination of topology and matter has sparked a tremendous surge of interest among theoreticians and experimentalists alike. Novel theoretical descriptions and predictions, as well as complex experimental setups confirming or refuting these theories, continuously appear in renowned journals. This review aims to provide conceptual tools to understand the fundamental concepts of this ever-growing field. Superconductivity and its historical development will serve as a second pillar alongside topological materials. While the primary focus will be on topological superconductors, other topological materials, such as topological insulators and topological semimetals, will also be explained phenomenologically. 
\end{abstract}

\begin{multicols}{2}
\section{Introduction} 
Topological quantum materials are a class of materials that exhibit topological properties in their electronic band structures. These materials have attracted significant attention in condensed matter physics due to their unique electronic properties and potential applications in quantum computing and electronics. Some aspects of Topological Materials (TMs) were already theoretically described as early as the 1970s in 1D and 2D Ising models of superconductors by Thouless and Kosterlitz in what is now known as Kosterlitz-Thouless phase transitions\cite{kosterlitz2016kosterlitz}. Eventually they shared the Physics Nobel Prize in 2016 with Haldane. In contrast to other types of materials, TMs often have an edge over other types of materials in regard to predictability: Larger size TMs have greater robustness against disorganisation of the periodic lattice than conventional ones. They can therefore maintain the material's electronic structure better\cite{wieder2021topological}. The topic of TMs is a vast and quickly growing field, as it promises novel phenomena, materials and many applications in industry alike. Novel materials include Topological Insulators (TIs), Magnetic Topological Insulators (MTIs), Topological Semimetals (TSs) and Chiral Crystals (CC)\cite{kane2005quantum}\cite{menth1969magnetic}\cite{crassee20183d}. While may consider the discovery of the quantized Hall Effect, later more often referred to as the Quantum Hall Effect (QHE) in 1980 by Klaus von Klitzing as a birthdate to topological states of matter, many theorists as well as experimentalists have revolutionized the field over and over again \cite{vonKlitzing1986quantized}\cite{bernevig2013topological}\cite{ando2015topological}. While some TMs such as TI have been extensively studied \cite{qi2011topological}, fields such as CC especially \cite{chang2018topological}(if not all) require further resources. As for additional literature, one is also referred to references \cite{yan2012topological} and \cite{liu2019topological}.

To show the already extensive family of TMs and phenomena that have emerged over the previous decades, figure \ref{firstdiagram} gives an overview of this exciting new subbranch of solid state physics, starting with the QHE in 1980 \cite{vonKlitzing1986quantized}. The effect had not been predicted theoretically \cite{Yoshioka2002}. The QHE as well as the Quantum Anomalous Hall Effect (QAHE, discovered in 1988)  are only given as two examples of the many topological phenomena known today \cite{pronin2021advances}. Other topological classes include the Quantum Spin Hall Effect (QSHE) \cite{spinhall} and the Fractional Quantum Hall Effect (FQHE) \cite{stormer1999fractional}.
\end{multicols}
\end{titlepage}

\begin{figure*}
    \centering
    \includegraphics[scale=0.85]{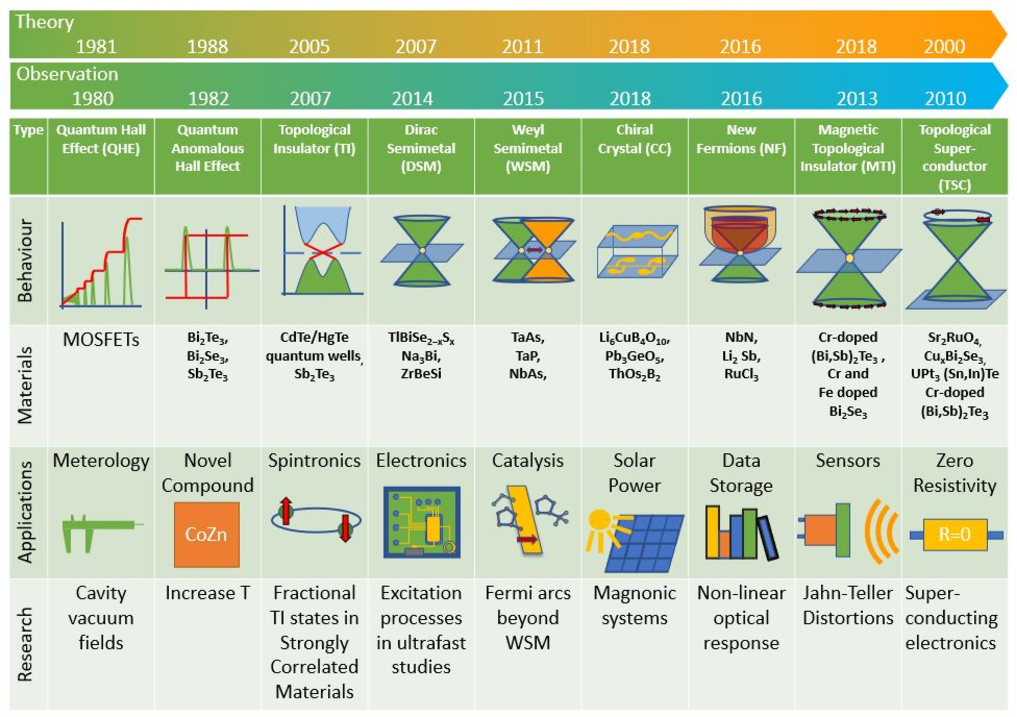}
    \caption{Topological Phenomena and Materials. In the top-most arrow-bar, the year of the paper of the theoretical prediction is given, followed by the experimental discovery (in topological terms). Additionally, the table gives insight into the currently most interesting characteristic topological phenomena or materials, including diagrams showing characteristic conductance diagrams (only applies to QHE and QAHE) or the energy dispersion relations, qualitatively. Furthermore, exemplatory applications and current topics of research are provided. Data taken from \cite{yu2010quantized}\cite{Landauquanti}\cite{armitage2018weyl}\cite{chang2018topological}\cite{konig2007quantum}\cite{chang2017nexus}\cite{wang2021intrinsic}\cite{ando2015topological}\cite{kane2005quantum}\cite{laughlin1981quantized}\cite{li2019exploring}\cite{weng2015weyl}\cite{chang2018topological}\cite{sau2010generic}\cite{appugliese2022breakdown}.}
    \label{firstdiagram}
\end{figure*}

\begin{multicols}{2}
The behaviour of materials and effects can be seen in the corresponding row "Behaviour" in figure \ref{firstdiagram}. The QHE effect follows a characteristic longitudinal and transverse resistivity $\rho_{xx}$ and $\rho_{xy}$ curves, discussed in more depth in a following chapter. Additionally, the characteristic diagram of the QAHE is shown as well with the $\rho_{xy}$ in red and $\rho_{xx}$ in green. Diverting the attention from the characteristic behaviour of the QHE and the QAHE, we then go on to show the characteristic electronic dispersions of different TMs, starting with the oldest of them: the TI with SOC. Furthermore, the 3D equivalent to the electronic structure of graphene is shown: a Dirac Semimetal with its two cone-like dispersions meeting at the Fermi level. From there, when certain symmetries are broken, it is possible to construct a Weyl Semimetal. The electronic properties of a Chiral Crystal are very different to previously described as shown in the schematic binding energy cuts of the electronic spectrum. Novel kinds of electronic structures with spin dependencies are shown for the New Fermions, Magnetic Topological Insulators and Topological Superconductors which are all rather recent developments in science.

On the materials side, a multitude of materials such as TIs, SMs, CCs, MTIs as well as specifically Topological Superconductors (TSCs) will be covered here, if only in a phenomenological and introductory manner.  We will later observe that many phenomena are often very similar to each other and hence also the possible applications. Of such character are the relationships between the QAHE with MTIs and QSHE with TIs. Even though the table of figure \ref{firstdiagram} suggest a strict categorization, especially applications may often overlap. TMs have not found their rightful place in engineering and electronics yet. However to construct smaller, more efficient and faster devices, it will be hard to dismiss promising topological technologies, an example being conventional field-effect transistors (FETs) being replaced by topological FETs \cite{gilbert2021topological}. Other promising fields of applications include meteorology \cite{poirier2009resistance}, novel compounds \cite{he2015quantum}, spintronics and electronics \cite{gilbert2021topological} \cite{he2022topological}, catalysis (reduction, adsorption, etc.)\cite{li2020heterogeneous}, data storage and conversion \cite{luo2022topological} as well as sensors \cite{budich2020non} and especially quantum computing applications for TSCs\cite{moore2010birth}. Figure \ref{firstdiagram} depicts the chronology of topology by means of categorization: Sr$_{2}$Ru0$_{4}$ was already discovered to be superconducting in 1994 \cite{maeno1994superconductivity}\cite{sato2017topological}, however it was only discovered much later to be of topological nature and named to be 'topological' in 2015 \cite{imai2015topological}. 

This review is divided into two main sections. First, we will address the fundamental concepts of topology in mathematics, introducing its application to materials science. We will then explore essential topological effects in materials, such as the QHE and the QSHE, acknowledging the wide array of topological effects known today \cite{spinhall}. Additionally, various TMs will be discussed in a phenomenological context. In the second section, we will delve into the history and development of superconductivity. Finally, we will examine these concepts in the chapter on TSCs, examining how topology in materials intersects with superconductivity.

\section{Fundamental Concepts of Topology}

Topological phases refer to unique states of matter characterized by the topological properties of their wavefunctions or energy spectra. These phases emerge in condensed matter physics and significantly impact material behavior. Topology provides a framework to classify whether two wave functions (such as those of electrons or quasi-particles) can be connected adiabatically, meaning through infinitesimally small, continuous changes. If one is able to transform a wave function adiabatically from one to another, these two objects (wave functions) belong to the same topological class (called $genus$) \cite{sato2017topological}\cite{ando2015topological}. As an example, a torus may be squeezed and deformed, however in order to be called a torus, it must have one (one hole specifically) hole at all times (remain in the same class of objects). Otherwise the classification of a torus no longer holds true. 

When examining possible new TMs, one is thus looking for invariant/breaking symmetries that best describe the macroscopic behaviour. Non-topological quantum materials typically have band structures that can be described by conventional band theory. In these materials, the energy bands and the electronic states within them can be continuously deformed without changing the topological properties. In contrast, topological quantum materials possess band structures with nontrivial topology, characterized by the presence of protected surface or edge states, nontrivial band crossings, or nontrivial topological invariants, such as the Chern number $C$ or $Z_{2}$ \cite{pollmann2015klassifizierung}\cite{kane2005quantum}. 

The Chern number is a prominent example to classify a topological invariant. It does not change under adiabatic evolution. It may be computed by taking the integral of Berry curvature over the Brillouin zone, where $\Omega_{xy}$ is the Berry curvature.

\begin{equation}
C=\frac{1}{2 \pi}\int_{BZ}\Omega_{xy}d^{2}k
\label{Chernnummer}
\end{equation}

The Berry curvature originates from a rotation (curl) of the Berry phase, which is a phase being picked up by a state under cyclic evolution. A quantum mechanical state that changes slowly enough ($=$ adiabatically) under a time-dependent Hamiltonian does not undergo sudden energy-Eigenstate transitions, but rather adjusts to the changing Hamiltonian. Additionally, the phase $e^{i \phi(t)}$ that is being picked up will not show in the expectation value of the Hamiltonian. Rather, the Berry phase becomes relevant upon superposition phenomena of two states \cite{rohrlich2009berry}.That is also why there is an ambiguity in the overall phase. One may choose it freely, however its relative phase to other state's phases is crucial (superposition). In that regard the Berry phase is very similar to the vector potential, as known in classical electrodynamics \cite{kane2013topological}. The Berry curvature in solids was recently experimentally detected by Schüler et al. by photoemission spectroscopy \cite{schuler2020local}.

In terms of TMs, if one wants to transition from one topological phase to a topologically distinct one, the energy gap in the energy dispersion relation must close at some point in the process. If this were not the case, the two phases could be adiabatically connected, thus be the same topological object \cite{kane2013topological}. Therefore a topological material (e.g. $C(Cu_{x}Bi_{2}Se_{3})=1$) that is positioned on a topologically trivial matter (e.g. air, $C=0$), the energy gap must close, which results in metallic surface states \cite{pollmann2015klassifizierung}. The Berry curvature may also be considered as entanglement between the conduction and valence band of the system \cite{yan2017topological}.

\subsection{Symmetries-Parity and Time Reversal}
Symmetries, particularly parity symmetry and time-reversal symmetry, are fundamental in understanding and characterizing topological quantum materials. These symmetries provide a basis for distinguishing different topological phases and predicting their properties. Before delving deeper into the topic, it is essential to explore these two key physical symmetries, as they hold special relevance in topology.

In condensed matter physics, physical systems are often classified based on the symmetries they preserve or break. For instance, topological insulators and superconductors exhibit unique properties depending on their symmetry classifications. The interplay between parity and time-reversal symmetries can lead to novel phenomena, such as the protection of edge states in topological insulators or the emergence of Majorana fermions in topological superconductors. Understanding these symmetries and their implications helps in the comprehensive characterization of topological quantum materials, guiding both theoretical predictions and experimental discoveries. As an example, crystals break translational symmetry \cite{maciejko2011quantum}. Thus, two symmetries will be discussed: Parity and time reversal. 

The parity operator acting on a wave function is often seen as a mirror in space, where $x \rightarrow -x$, $y \rightarrow -y$ and $z \rightarrow -z$ \cite{henley1969parity}. If a quantum-mechanical system (such as the electronic structure) remains unchanged under a parity transformation, the system is considered to be "even" parity. If, under a parity transformation, the system reverses its sign (negative to positive or vice versa), it is considered "odd" parity. The total angular momentum $J$ for a spinless free particle would be such a candidate for even parity, whereas a free fermion has odd parity \cite{fano1996symmetries}. The fermionic parity operator tells us, if we have an odd or even number of fermions we are working with \cite{turner2011topological}.

Time-reversal symmetry is a fundamental symmetry that relates the forward and backward evolution of a physical system. It states that if a system evolves from an initial state to a final state, the time-reversed evolution should also be allowed. In the context of topological quantum materials, time-reversal symmetry plays a central role in protecting and characterizing various topological phases \cite{henley1969parity}. Speaking in terms of classical mechanics, entropy would be TRS breaking, except if in equilibrium. In quantum mechanics, spins flip under a time-reversal operator, thus often (not always as we will see) leading to TRS breaking. The cause of this spin flip may be made in a semi-classical approach: Any magnetic moment flips under a time-reversal operator, thus also spins \cite{Spinfliptime}. The QHE is a good example for time-reversal-symmetry (TRS) breaking, while QSHE is prominent for time-reversal-invariance (TRI). Beyond the phenomenological approach, the time reversal operator applied upon a wave function results in the the complex conjugate of it \cite{bernevig2013topological}.

Even though symmetry breaking schemes at first might sound rather non-intuitive and merely theoretical, Bernevig, Hughes and Zhang were able to demonstrate the time-reversal invariance in HgTe/cdTe quantum wells as early as 2006 and the quantum Hall effect (despite not knowing the fundamental topological mechanisms then) has been around since the 1980s \cite{vonKlitzing1986quantized}\cite{bernevig2006quantum}\cite{thouless1982quantized}.

\subsection{Majorana Zero Modes}

Majorana zero modes (MZMs) are quasiparticles that have attracted significant attention in the field of topological quantum materials, particularly in the context of topological superconductors. A Majorana zero mode refers to a quasiparticle excitation that is its own antiparticle. MZMs are fermionic modes that have zero energy (hence "zero mode") and exhibit nontrivial braiding statistics. Unlike conventional fermions, which can be either occupied or unoccupied, MZMs are regarded as a superposition of electrons and holes \cite{majorana1937teoria}\cite{li2019exploring}. Neutral excitations in the bulk superconductor may then be identified as Majorana fermions. It is here where many TMs exhibit interesting (surface) phenomena \cite{li2019exploring}.

Several experimental techniques have been employed to detect and probe MZMs in topological quantum materials. One common approach is tunneling spectroscopy using scanning tunneling microscopy (STM)\cite{jack2021detecting}, which can reveal the presence of localized zero-bias conductance peaks associated with MZMs. Other techniques, such as Josephson junctions and interferometry, have also been utilized to study MZMs and their properties \cite{black2011majorana}\cite{akhmerov2009electrically}. 

\section{Hall Effects}
Equipped with a foundational understanding of topology, we can now delve into the fascinating realm of various Hall-related effects. These phenomena bridge the theoretical framework of topology with experimental materials, providing concrete examples of topological principles in action.

The journey begins with the classical Hall Effect (HE), discovered by Edwin Hall in 1879. The Hall Effect describes the deflection of electrons under the influence of an electric field due to the Lorentz force when a perpendicular magnetic field is applied, resulting in the generation of a Hall voltage \cite{hall1879new}. This discovery laid the groundwork for an entire field of study, becoming a fundamental concept in condensed matter physics.

Since then, the Hall Effect has evolved into a cornerstone of modern research, leading to the discovery of several related phenomena that showcase the significance of topology in real materials. The Quantum Hall Effect (QHE) demonstrates quantized Hall conductance in two-dimensional electron systems subjected to low temperatures and strong magnetic fields. The Quantum Anomalous Hall Effect (QAHE) occurs in certain magnetic materials without an external magnetic field, driven by intrinsic magnetic order and strong spin-orbit coupling. The QSHE involves the generation of spin-polarized edge states in topological insulators, where spin currents flow without dissipation. Lastly, the FQHE reveals the presence of quasiparticles with fractional charge and statistics in two-dimensional electron systems under extreme conditions.

These Hall-related effects not only highlight the quantized nature of electronic transport but also underscore the profound impact of topology in condensed matter physics. By examining these phenomena, we gain insight into how topological properties manifest in experimental systems, paving the way for future advancements in the study and application of topological quantum materials.


\subsection{Quantum Hall Effect (QHE)}

A century after the discovery of the fundamental HE, 1982 brought a breakthrough in condensed matter physics by Klaus von Klitzing et. al. with the discovery of the QHE which fundamentally shifted the understanding of the HE. It was discovered that at sufficiently low temperatures and high magnetic fields the Hall resistivity of a material was quantised. Thus, a first topological invariant necessary to describe Hall resistivity in the QHE had been discovered\cite{thouless1982quantized}. His findings also marked the beginning of TMs. He was awarded the Physics Nobel Prize in 1985\cite{vonKlitzing1986quantized}. 

The resistivity of the quantum Hall effect may be computed as follows, where $R_{Hall}$, $h$, $C$ and $e$ are the Hall resistance, the Planck constant, an integer and the electron charge respectively:

\begin{equation}
    \rho_{Hall}=\frac{h}{Ce^{2}}
\label{QHEequ}
\end{equation}

From this equation alone it becomes clear that the resistivity (and thus conductivity) is quantized by an integer number $C$, the Chern number.

\begin{figure}
    \centering
    \includegraphics[width=1\textwidth]{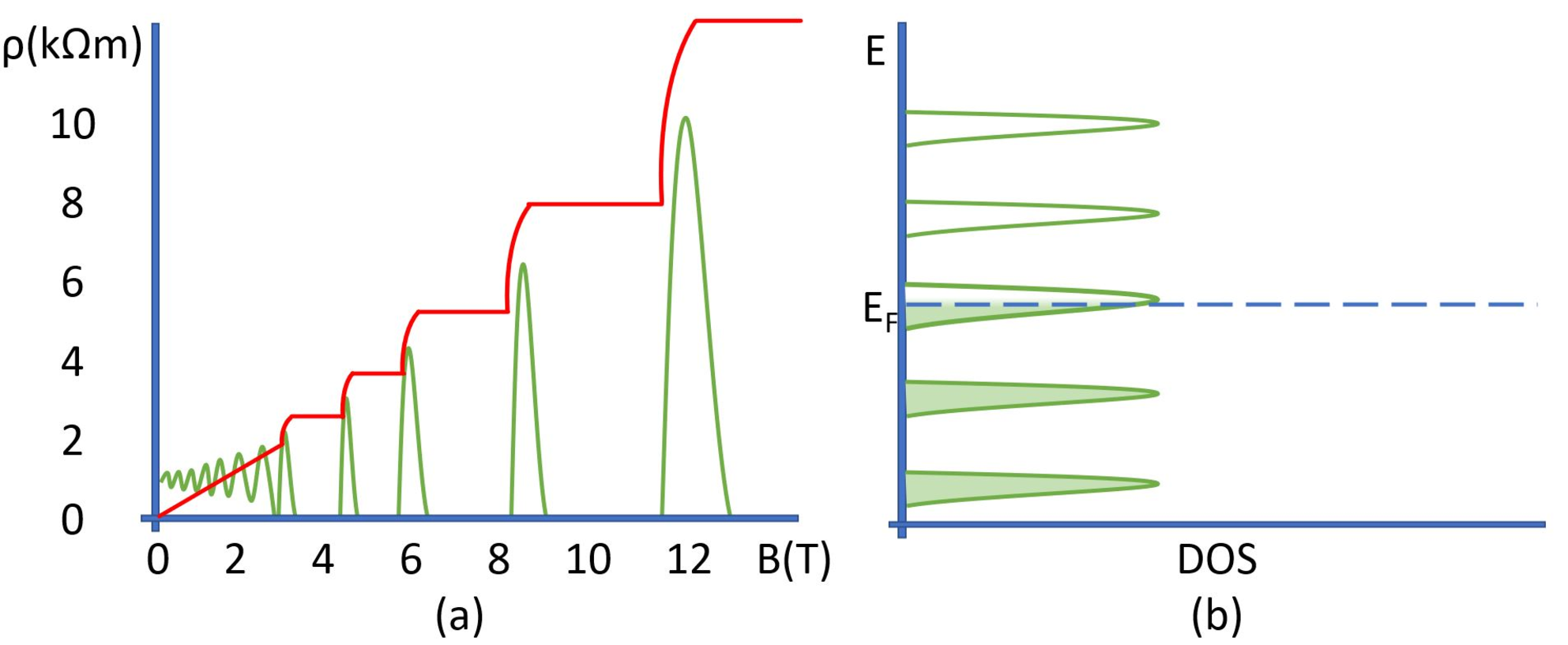}
    \caption{Qualitative characteristic functions of the QHE dependent on magnetic field strength B. (a)Schematic characteristic functions of the transverse (red) and longitudinal (green) resistivity.   (b) Quantized Landau levels under a high magnetic field B. Adapted from \cite{vonKlitzing1986quantized} and\cite{Landauquanti}}
    \label{QHEdiagram}
\end{figure}

To get a better feeling of the effect and how the Chern number may be determined through experiment, we will look into figure \ref{QHEdiagram}. The two subfigures explain the QHE by the characteristic functions (red and green) as well show the typical DOS at high magnetic fields derived by Landau.

Before we will investigate what happens in the schematic characteristic function of the QHE, we must first understand Landau levels, figure \ref{QHEdiagram} b). Quantised Landau levels arise due to large magnetic fields (magnitude of $T$) that squeeze the density of states (DOS) and may be derived by the Hamiltonian of a 2D film with a perpendicular magnetic field acting upon it \cite{landau1930diamagnetismus}.

\end{multicols}

\begin{multicols}{2}

The separation between the levels is proportional to the cyclotron frequency $\omega_{c}$, where $q$, $m$ and $B$ correspond to the charge and mass of the electron as well as the applied magnetic field:

\begin{equation}
    \omega_{c}=\frac{qB}{m}
\end{equation}

One realizes instantly that the distance between the Landau levels shrinks in decreasing $B$. Additionally, Landau levels broaden due to electron-electron interaction and impurities of the system \cite{Landaubroadening}.  When the Fermi energy (denoted as $E_{F}$) crosses a Landau level, both longitudinal (green in figure \ref{QHEdiagram} (a))  and transverse (red) resistivity decrease instantaneously. The broadened Landau level acts both as conduction and valence band, thus allowing current to flow and resistivity to drop. By further decreasing $B$, this Landau level shifts further down until it is no longer at the Fermi level, resulting in fully insulating materials. At low $B$, Landau levels become indistinguishable from each other due to their proximity and are positioned beneath the Fermi level. The longitudinal resitivity occurs in true quantisation, while the transverse resistivity occurs in plateaus, due to impurities of the system\cite{vonKlitzing1986quantized}.

A more detailed explanation of the QHE and Landau levels may be found in references \cite{vonKlitzing1986quantized},\cite{Yoshioka2002} and \cite{reis2013fundamentals}. Today, the QHE is applied in fields such as meteorology \cite{poirier2009resistance}.

\subsection{Quantum Anomalous Hall \\  Effect (QAHE)}

One might pose the question whether the QHE can also be induced without strong outer magnetic fields. Such behaviour may indeed be induced by an appropiate crystal structure and  chemical composition. Such an effect is known as the Quantum Anomalous Hall Effect (QAHE). 

Building upon the concept of the QHE-but by no means as famous- the QAHE is the quantized equivalent to the Anomalous Hall Effect (AHE) and closely related to MTIs. In the AHE, the deflection of electrons in a material may occur, even if no outer (or inner) magnetic field is present. The underlying reason for this non-quantized anomaly is ferromagnetism, SOC and disorders, as proposed by Yu et. al \cite{yu2010quantized} \cite{liu2015quantum}.

Like the QHE, the QAHE arises from a nonzero Chern number. The nonzero Chern number (a result from TRS breaking) may however not stem from a magnetic field which would again lead to Landau levels. As shown in equation \ref{Chernnummer}, to obtain the Chern number, one shall integrate over the Brillouin zone and its occupied bands. Due to other contributions, metals however will not exhibit integer Chern numbers and have trivial topology. Few materials fulfill all recquirements (also called Chern insulators) \cite{liu2015quantum}, however some material families exist:  

\end{multicols}
\begin{figure}[ht]
    \centering
    \includegraphics[width=1\textwidth]{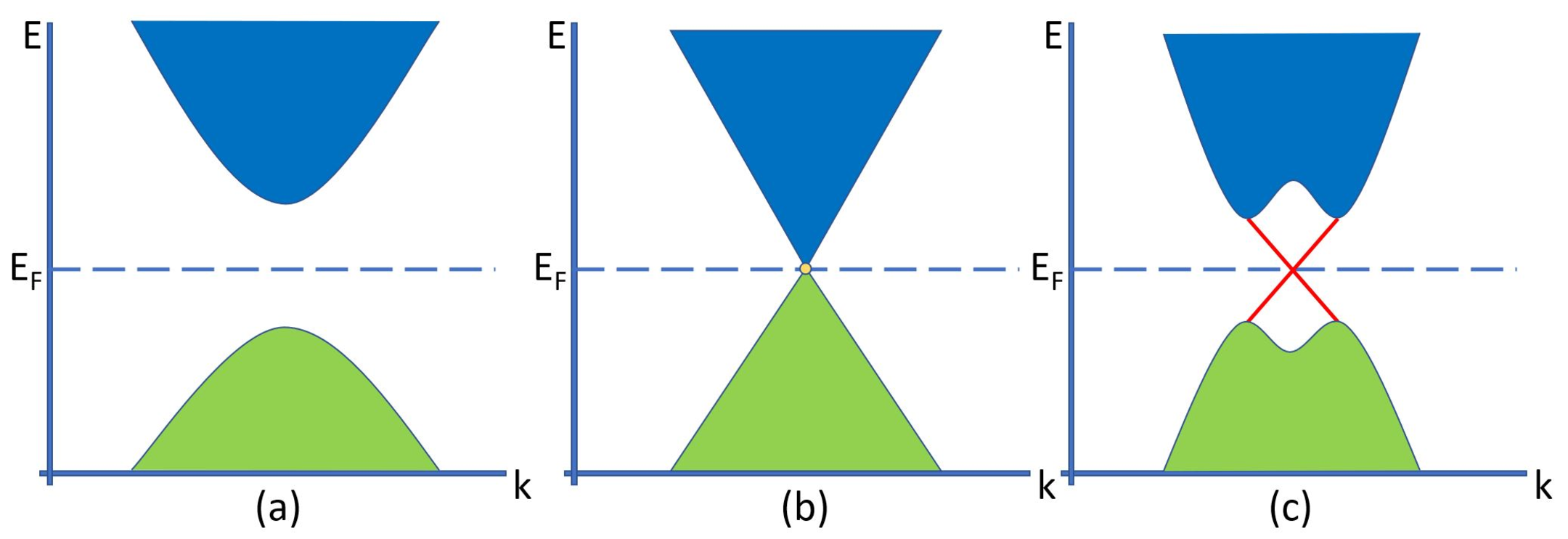}
    \caption{Schematic band structures of (green=valence band, blue=conduction band, red=surface states) of a regular insulator (a), a Dirac cone with a Dirac point (in yellow) (b) and a TI (c). The TI exhibits surface states, indicated in red as well as band inversion due to strong SOC.}
    \label{Diraccones}
\end{figure}

\begin{multicols}{2}

Magnetically doped topological insulators like (Bi,Sb)$_{2}$Te$_{3}$, thin films of the magnetic topological insulator $MnBi_{2}Te_{4}$, topological semimetals \cite{crassee20183d} and moiré materials formed from graphene and transition metal dichalcogenides \cite{chang2016quantum}.

\subsection{Quantum Spin Hall Effect \\  (QSHE)}

The QHE recquires a magnetic field to induce a quantum Hall state (QHS). While the implications of such a system are remarkable, the applied magnetic field breaks TRS of the system. Soon after its discovery, scientists started working on a spin-channel seperated equivalent of QHE, which would allow TRS to remain intact. However, it remained unclear, whether one could realistically separate the spin current channels and tackle possible impurities\cite{moore2010birth}. 

In 2005 however, equipped with a new topological invariant $Z_{2}$ (comparable to the Chern number, but suited for TRI systems), Kane and Mele predicted that the QSHE in graphene would keep TRI, but also remain in a QHS\cite{kane2005quantum}\cite{kane2005z}. They took the crucial step of splitting up the Hamiltonian of the QHS into two seperate ones. That resulted in two identical copies of the Hamiltonian but with opposite spins (up and down). Thus, they took into account the spin orientation of the electrons. The overall system will remain in its original QHS because upon the time reversal operation, the two Hamiltonians will flip their spin orientations but result in each other \cite{kane2005quantum}.

 In two dimensions, there exist two resulting edge states, one for each spin channel. No net current will flow if an electric field is applied, since the electrons of different spin channels move in opposite directions. Net spin currents however do flow\cite{kane2005quantum}. These theoretical predictions were soon confirmed experimentally in 2006 by Bernevig et al. in HgTe\cite{bernevig2006quantum}.

\subsection{Fractional Quantum Hall\\ Effect  (FQHE)}

Yet another contribution to the canon of Hall related effects was made by Robert Laughlin, Horst Störmer, and Daniel Tsui. They received the Nobel prize in 1998 for their work on a fractional quantization of the Hall conductance\cite{schwarzschild1998physics}. They discovered, that the integer plateaus, as shown in figure \ref{QHEdiagram}, may also take fractional values in 2D electron gases that condensated into a 2D quantum fluid\cite{henini2012molecular}. The FQHE primarily arises in strongly correlated systems\cite{laughlin1983anomalous}\cite{liu2015quantum}.  Consequently, equation \ref{QHEequ} must allow for non-integer charges. In a nutshell, this may be explained by a quasiparticle (instead of electronic) excitation, making it neccessary to introduce fractional (going beyond Bose and Fermi) statistics to the problem \cite{feldman2021fractional}. An example of material (due to its high mobility) is GaAs and graphene\cite{lin2014recent}. Generally, it may also be observed in WSM \cite{wang2020fractional}.

\section{Topological Materials}

Now that the understanding of Hall-related effects is well established, TMs themselves will be the main focus of this chapter. It will start with perhaps the most famous of TMs: TI. Then, TSMs shall be studied a bit more in depth, focusing on DSM and WSM. Finally, CC and MTI will be tackled.

\subsection{Topological Insulator (TI)}

To better understand the pyhsical consequences of the Chern number and of topology in general, we shall investigate an arbitraty topological insulator: In a regular insulator the band gap between the valence and conduction bands is too large, hence no electrons are excited into the conduction band (readily described by classical band theory). However a topological insulator also displays interesting phenomena on the surface. These so called surface states/bands can exhibit conducting, superconducting, magnetic, antiferromagnetic behaviour, etc. That is what makes topological quantum materials different from trivial materials \cite{kane2005quantum}. These surface bands may often be characterized by Dirac cones, which may become clearer with figure \ref{Diraccones}. These surface bands in the energy dispersion relation closely link to symmetry protected states as discussed with topology.

As can be seen in figure \ref{Diraccones} a), a trivial band structure of an arbitrary insulator is shown. Figure \ref{Diraccones} $(c)$ displays a typical band structure of a topological insulator. Using classical band theory, the valence band and conduction band should readily overlap resulting in a conductor. However, due to so called band inversion, a very special band structure arises. Band inversion stems from the Spin-Orbit-Coupling (SOC) of the system. SOC describes the magnetic field induced by the moving atomic nucleus acting on the electron (and its spin-state) in the reference frame of the electron \cite{thomas1926motion}. While the band gap between the two bulk phases of the insulator are still too wide for any electrons to transition between the valence and conduction band, conducting surface states occur (red). The first experimental discovery of such surface states was made in 2008 by Hasan et al. \cite{hsieh2008topological}. As already mentioned, the underlying theoretical concepts were developed by Kane and Mele in terms of the Quantum Spin hall effect \cite{kane2005quantum}. 

Figure \ref{Diraccones} $(b)$ depicts a DSM. One may see the DSM as an intermediate material between the trivial insulator and the topological insulator. DSM have gapless energy bands and are topologically as well as symmetrically protected \cite{armitage2018weyl}. DSM will be tackled in the next chapter.

Over the previous years, yet another subdivision of TMs research has emerged: Topological Kondo Insulators (TKI)\cite{dzero2016topologicalkondo}. This sub-field brought together two fields of physics: Heavy fermion and topological systems. The Kondo effect itself was already discovered in the 1930s, describing the electron-electron interaction which is stronger than the kinetic energy of the electrons in heavy fermion systems. That in turn gives rise to high electron-electron correlations and exotic phenomena. There exists a characteristic Kondo temperature $T_{K}$ (not to be mistaken by the critical temperature $T_{C}$ of SCs) above which the material behaves like a metal with localized magnetic moments (also referred to as magnetic impurities in the lattice), because electrons only scatter weakly off those impurities at high temperatures. Below $T_{K}$, scattering becomes a dominant phenomenon, leading to the creation of spin singlets between the magnetic disorders and electrons. That in turn leads to distinctive band gaps in the energy dispersion relation\cite{dzero2016topologicalkondo}. The first TKI to be discovered was SmB$_{6}$ in the late 1960s\cite{menth1969magnetic}\cite{dzero2016topologicalkondo}.

\subsection{Topological Semimetals \\  (TSMs)}
Another interesting exotic topological phenomenon are TSMs. Amongst the most famous are DSM and WSM\cite{crassee20183d}. In theory, a distinction between the two may be considered almost trivial: Electrons in DSM follow the Dirac equation \cite{dirac1928quantum}, while electrons in WSM follow the Weyl equation \cite{weyl1929electron}. Phenomenologically, they however differ substantially. Both Weyl and Dirac fermions assume vanishing effective masses of the fermion (in our case the electrons in the crystal) \cite{dirac1928quantum}\cite{weyl1929electron}. Opposed to TI, topological semimetals already have bulk conductivity but may exhibit similar or identical metallic topological surface states\cite{crassee20183d}.

\subsubsection{Dirac Semimetal (DSM)}

\begin{figure}[ht]
    \centering
    \includegraphics[width=1\textwidth]{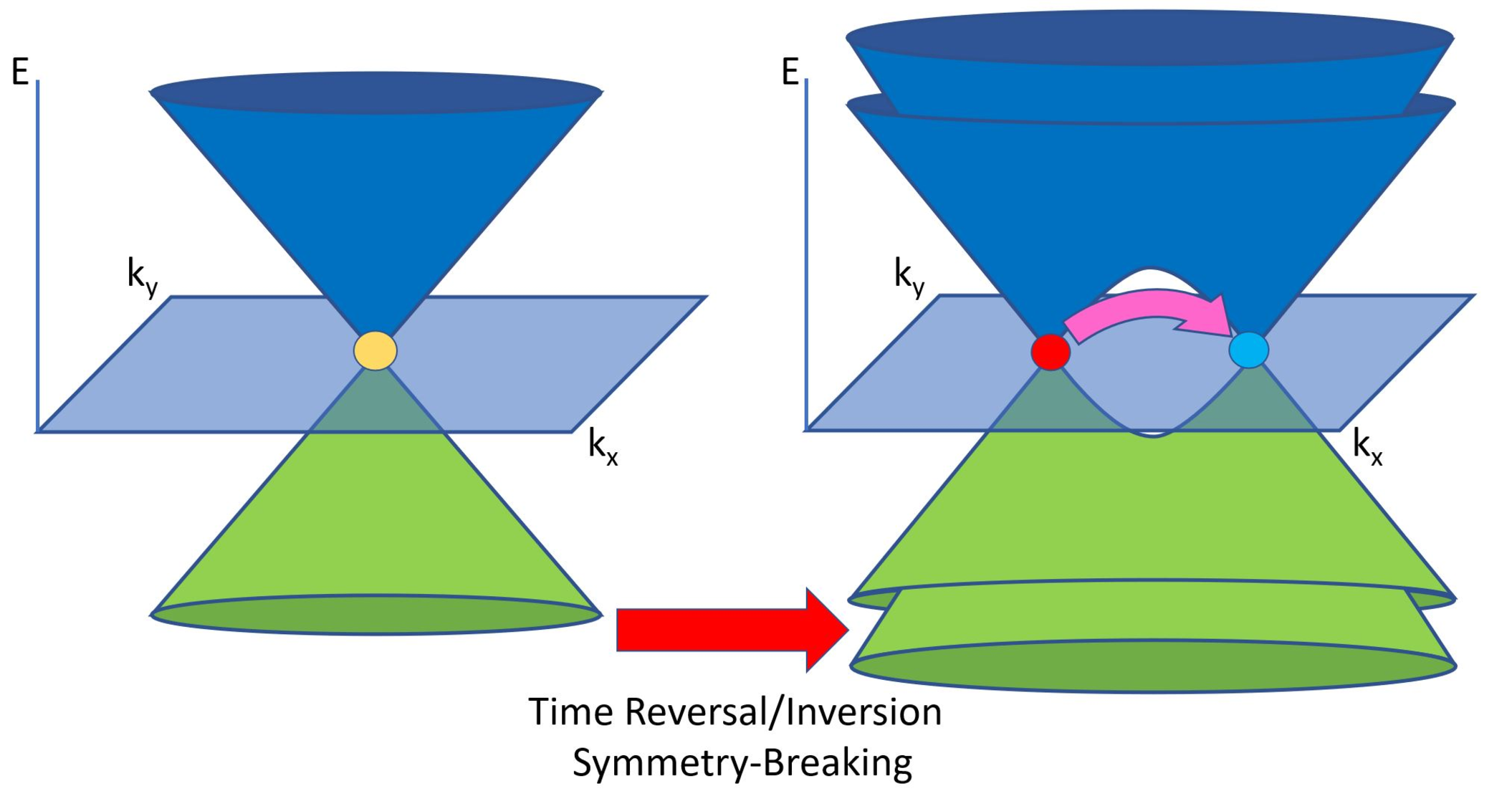}
    \caption{Three dimensional schematic band structures of a) DSM and b) WSM. In DSM the conduction (blue) and valence (green) bands are connected at a Dirac point (yellow). When TRS or inversion symmetry is broken, the Dirac point splits up into two Weyl points of opposite chirality, which are connected at the surface through a Fermi arc (pink). Adapted from \cite{Lau_phdthesis}.}
    \label{Diracweyl}
\end{figure}

A DSM-like band structure is composed of two Dirac cones, one of them being inverted and connected by a so called Dirac point, which is ideally placed as close as possible to the Fermi level, as shown in figure \ref{Diracweyl}. 

\end{multicols}
\begin{multicols}{2}

The energy dispersion in the band structure thus exhibits 3D linear (conical) behaviour very much like the 3D equivalent to graphene \cite{zhang2005experimental}\cite{yan2017topological}. Graphene does however not have any topologically protected surface states. If put on a substrate, Graphene's  energy gap opens up. Due to topological protection, that would not happen to a DSM. 

Like in TI, spin and momentum are also locked up in topological semimetals. Oftentimes (such as in $Cd_{3}As_{2}$), the conical band structure arises due to the strong dependence of the effective mass of the system on the electron concentration and SOC\cite{crassee20183d}. 

While research in the example of $Cd_{3}As_{2}$ had already started in the 1930s as the interest of scientists shifted to semiconductors, it was only in 2013, that Wang et al. \cite{3dCDAS_Wang} predicted Dirac cones in materials\cite{crassee20183d}. Previously, other models like the Kane model on semiconductors described the material quite well via $k\cdot p$ perturbation theory \cite{KANE1957249}. If the band gap is small compared to the energy scale of the material, conical structures like in DSM in the band structure arise.  However, these models did not incorporate the essential protected surface states. Wang's prediction was already verified a year later by Rosenman et al. \cite{ROSENMAN19691385}. 

\subsubsection{Weyl Semimetal (WSM)}
WSM have received favorable attention in the previous years as quantum computing and spin transport applications are to be expected in the future. WSM superconductors already exist \cite{yan2017topological}. Similarly to TI, band inversion occurs in WSM, as may be seen in figure \ref{Diracweyl}, when TRS or inversion symmetry are being broken in a DSM.

When either of those symmetries are broken, the Dirac point seizes to exist/is split up and from now on two doubly degenerate Weyl points \cite{li2019chiral} of opposite chirality (positive or negative such as in dipoles) emerge. The Berry curvature at these points is either positive (source) or negative (sink) in chirality. This makes an uneven number of Weyl points impossible, since otherwise the Berry curvature would be confronted with divergence issues \cite{yan2017topological}.  There exists however one exception: When TRS and inversion symmetry endure, the Weyl points become degenerate in a single point, once again a Dirac point \cite{yan2017topological}. Thus by tuning the symmetries one may transition from one topological semimetal into the other. Experimentally, that is not so easily done \cite{3dCDAS_Wang}\cite{yan2017topological}. Finally, it is usually necessary to bring the Fermi level as close as possible to the Weyl points. That may be achieved via electron doping \cite{sergelius2016berry}. 
\end{multicols}

\begin{figure}[ht]
    \centering
    \includegraphics[width=1\textwidth]{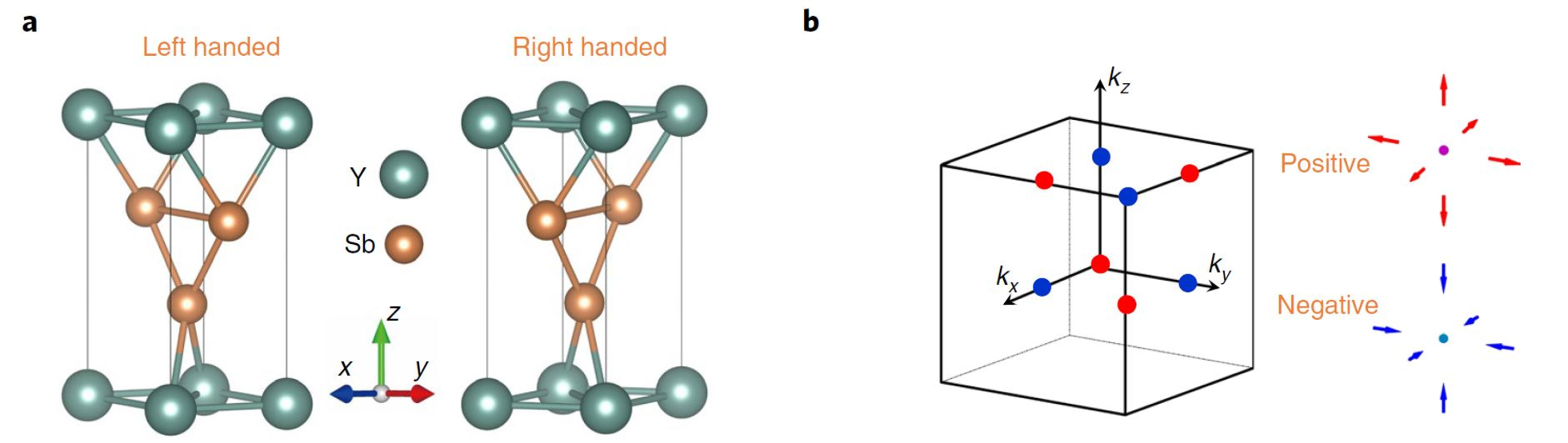}
    \caption{Chiral Crystals. a) Example of inversion and rotation symmetry breaking leading to right and left handed materials. b) Chiral fermions responsible for topologically non-trivial Chern numbers. Taken from \cite{chang2018topological}.}
    \label{Chiralcrystaldia}
\end{figure}
\begin{multicols}{2}

In short, Weyl points, closely connected to the gapless surface states at the Fermi level are protected by surface topology and are of opposite chirality. Opposite chirality Weyl points are connected by Fermi arcs, which are metallic surface states \cite{xu2015discovery}.

Current research has recently taken a special interest in type-II WSM. This advancement of the original WSM introduces a tilt of the Weyl cone with respect to the Fermi plane and induces the QAHE. The QAHE is caused by a non-zero Chern number that in turn results from the Berry phase that is being picked up in the $k_{x}$-$k_{y}$-plane. Finally, it is noteworthy to mention that interesting phenomena like negative magnetoresistance under parallel magnetic and electric fields and exotic transport properties which arise due to the chiral anomaly in WSM emerge\cite{xu2015discovery}\cite{nielsen1983adler}.

\subsection{Chiral Crystal (CC)}

Materials that do not exhibit any rotational or inversion symmetries (including a combination of both called roto-inversion) may often be classified as CCs which is in contrast to other materials that have been introduced here. Such lack of symmetry often leads to a topological band structure \cite{chang2018topological}.

CC are interesting, because of their exotic optical and magnetic responses. Chiral magnets for example could be used to induce skyrmions \cite{bogdanov1994thermodynamically}\cite{chang2018topological}. A material of the CC family is said to possess a certain handedness with regards to its crystal structure \cite{chang2018topological}. Such handedness arises from the massless Weyl-fermion (the massless electron following the Weyl equation\cite{weyl1929electron}).

The fermion shows certain spin-behaviour (up or down), which cannot be mirrored, rotated or inverted. Thus, the spin up (or down) of the electron results in the handedness. This is shown in figure \ref{Chiralcrystaldia} a), where the given crystal structure cannot be mirrored, rotated, or inverted.

To understand CCs, a crucial concept is points of time-reversal-invariant momenta (TRIM). Such TRIM points may be considered nodal fermions which sit at specific spots in the Brillouin zone and are the driving effect for various phenomena in CC. Chang et al. \cite{chang2018topological} showed that nodes with non-trivial Chern numbers arise from Kramers degeneracies at TRIMs. Kramers degeneracies occur in systems of half-integer total spin. Kramer's theorem states that there is minimally one other eigenstate of the same energy upon a time-reversal operation of the Hamiltonian if the system's total spin is a half-integer value\cite{kramers1930theorie}. They found chiral fermions (electrons) of opposite chirality in orthorhombic crystal symmetries, as shown in figure \ref{Chiralcrystaldia} b) \cite{chang2018topological}.

Combining CCs with the concept of semimetals, yielding chiral semimetals such as PdGa, promises yet another interesting path, which may be taken in terms of TMs \cite{sessi2020handedness}.

\subsection{Magnetic Topological Insulator (MTI)}

Essentially, MTIs have to fulfill some fundamental criteria: Ferromagnetic ordering and SOC must be strong enough for a non-zero Chern number to arise (to be topologically non-trivial). TR symmetry breaking is also required which is achieved by the insertion of (anti-) ferromagnetism into the TI. By such an introduction, gaps open up due to magnetic

\begin{figure}[ht]
    \centering
    \includegraphics[width=0.7\textwidth]{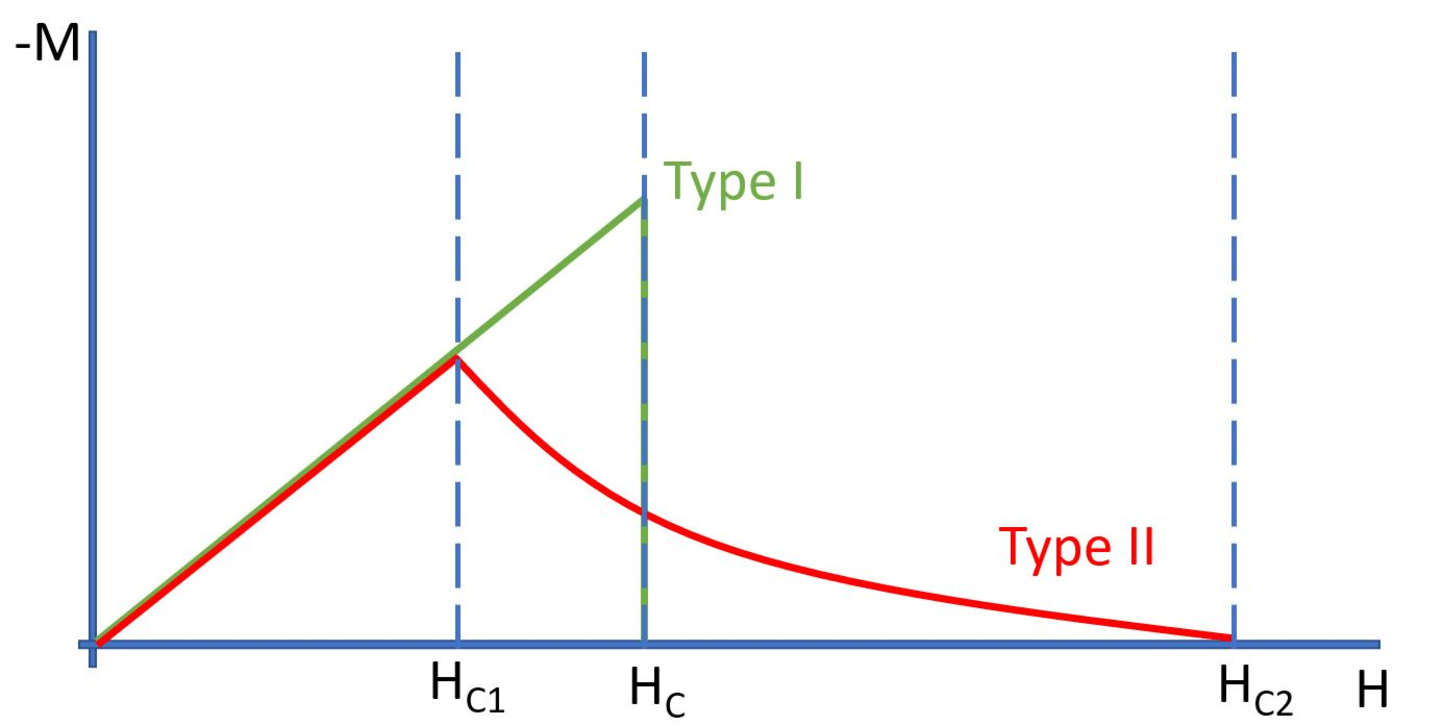}
    \caption{Schematic magnetization curves of type I and type II SCs. Adopted from \cite{magnetization}.}
    \label{magnetization12}
\end{figure}

exchange interactions\cite{wang2021intrinsic}.  Naturally, MTI exhibit QAHE behaviour \cite{wang2015quantum} and homogeneous magnetic ordering on the surface \cite{wang2021intrinsic}.

The history of the QAHE is closely related to the discoveries connected to the class of MTIs \cite{jiang2022review}. Initially, researchers were unsuccessful at finding suitable materials exhibiting the QAHE, since oftentimes extreme conditions were necessary. Another reason was that bulk bands were often mixing with the topological surface states, making numerous materials subsequently unfitting. By doping $(Bi,Sb)_{2}Te_{3}$ with $Cr$, Chang and associates were the first ones to detect the QAHE within a real material. Unfortunately, large magnetic fields as well as temperatures beneath $0.1K$ were necessary \cite{chang2013experimental}. With an increasing quality of synthetisation over time, temperatures of up to $2K$ became possible. In addition to doping, stacking TI and ferromagnetic heterostructures or introducing magnetic layers in TI  may also lead to MTI \cite{jiang2022review}.

\section{Superconductivity}

Since Heike Kamerlingh Onnes' ability to liquefy helium in 1908 and his subsequent discovery of superconductivity some three years later \cite{early}, great advancements in the field of superconductivity have been made and new research fields such as Topological Superconductors have emerged. This chapter shall investigate the early historical developments of superconductivity and will elaborate on primary milestones of this ever growing sub-discipline of condensed matter physics and the relatively recent theoretical approaches and experiments conducted specifically in the field of TSCs. 

\subsection{Early Experimental Findings}
After the Dutchman Heike Kamerlingh Onnes had accomplished sufficient cooling of liquefied helium (about 4.14 K) in 1908, he was then interested in resistivity behaviour of pure metals near absolute zero temperature, introduced by William Thomson (Lord Kelvin). According to Kelvin, resisitivity of any pure metal should linearly decline with temperature, but peak towards infinity at absolute zero. Kamerlingh Onnes first also supported that theory. Previous cooling techniques had however only allowed temperatures as low as 14 K, such that with Kamerlingh Onnes' new cooling technique further experimental progress was possible. He soon found that Kelvin's theory only holds for pure metals such as gold, platinum and mercury above certain material-dependent threshold temperatures. Resistivity plummits below such temperatures and becomes almost unaffected by a change in temperature by going even lower \cite{early} . He thus observed the phenomenon of superconductivity for the first time in human history.

Kamerlingh Onnes received the Nobel Price in Physics in 1913 \cite{nobel} promtly after his discovery in 1911. Kamerlingh Onnes also reported that relatively small magnetic fields would break the superconducting state \cite{year100}. His discovery would spur solid state physicists up until today, since the promise of his discovery was quite clear: Loss-free electron transfer across large distances.

Relatively few advancements in the field were made in the following two decades, even though superconductivity in alloys of two non-superconducting materials was discovered by W.J. De Haas and associates in 1928 \cite{alloy}. Yet another contribution to the growing field of superconductivity research was the experimental discovery of Type II-SCs by Shubnikov et al. in 1937 \cite{typ2}. This new type of SC had a very interesting property: While Type-I SCs' (all previously discovered SCs) threshold temperature decreased in an increasing magnetic field, Type-II SCs exposed peculiar behaviour. While these new superconducting materials behave like type-I SCs beneath a certain magnetic field strength, their superconducting state does not collapse above such a magnetic field. Rather, the magnetization decreases exponentially as can be seen in figure \ref{magnetization12}.

\subsection{Meissner-Ochsenfeld effect}
The next great breakthrough however came some years later: The Meissner-Ochsenfeld effect. Walther Meißner and Robert Ochsenfeld found that SCs will only allow a given magnetic field acting upon it to enter to a certain depth \cite{MOeffect} (later to be known as the London penetration depth). Meissner and Ochsenfeld found that this maximal penetration depth allows magnets to float above (and beneath) SCs \cite{MOeffect}. 

While one may look at Meissner and Ochsenfeld's discovery simply as one in a chain of physics research, one shall not underestimate the importance of their findings. For that reason we shall focus on the difference between a SC and a perfect conductor. 

As the name suggests the perfect conductor becomes perfectly conducting as the temperature falls beneath its critical value ($T_{c}$). However, the perfect conductor still follows Faraday's law:

\begin{equation}
    \oint E ds =-\frac{d\phi}{dt}
\end{equation}

$E$ denotes the electric field, $\phi$ is magnetic flux and $s$ is the integral path. The law states that a changing magnetic field through a surface $\phi$ with respect to time ($\frac{d\phi}{dt}$ is equal to the closed path integral of the surface boundary of the electric field $E$. Thus, when the magnetic flux changes, a voltage may be measured along the surface boundary (for example in a wire).

\begin{figure*}[ht]
    \centering
    \includegraphics[width=0.8\textwidth]{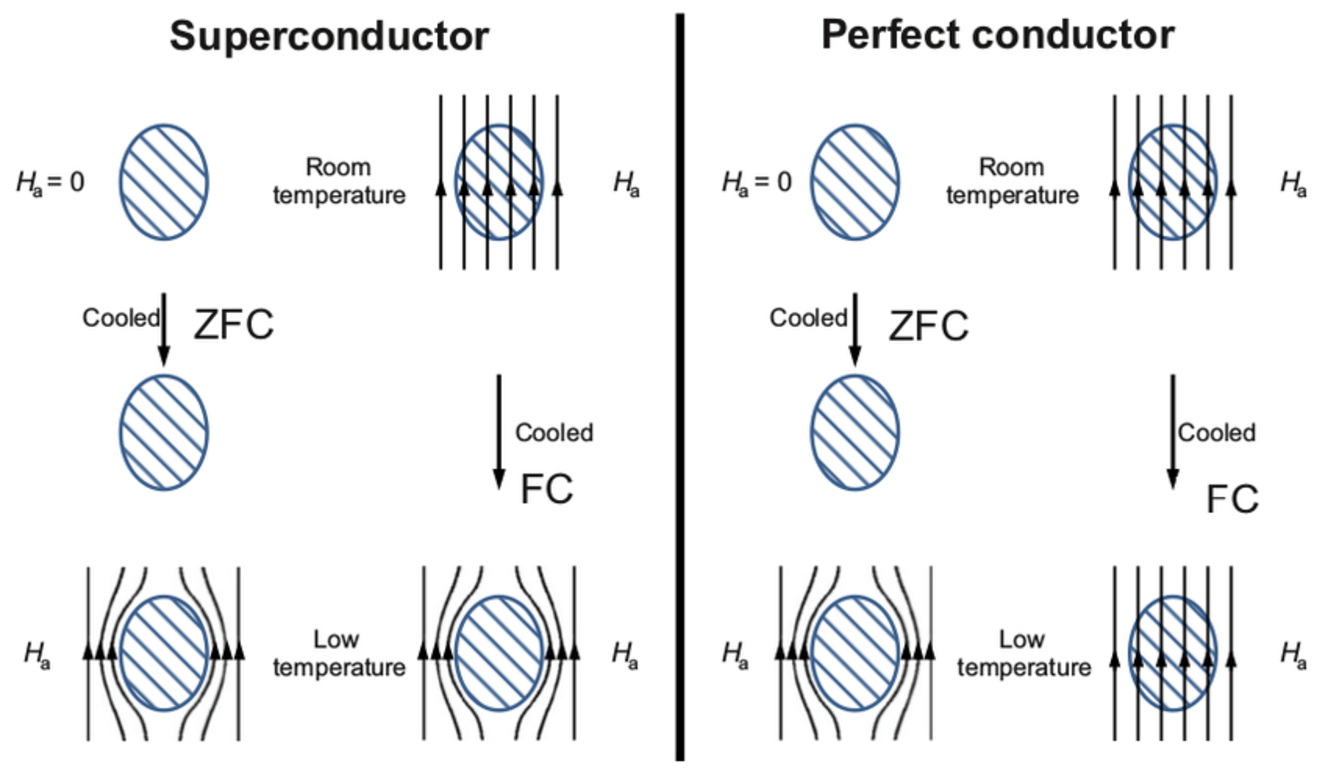}
    \caption{Schematic comparison of perfect conductor and SC. Taken from \cite{phdthesisrettaroli}.}
    \label{perfectconductor}
\end{figure*}

To get a better understanding of the particuliarity of SCs and the Meissner-Ochsenfeld effect specifically, we shall now study the difference between a perfect conductor and a SC in more depth. We shall begin with a SC and a perfect conductor at room temperature, with and without an applied magnetic field $B$ (see figure \ref{perfectconductor}). 

The perfect conductor behaves classically. To explain the perfect conductor, one should start with Maxwell's equations and that the response to any external electric field in the perfect conductor is instant (recquires no time). Thus, given an external electric field, an internal electric field arises simultaneoulsy with the external electric field, hence cancelling out the overall electrical field in the conductor. That yields $\Vec{\nabla} D=0$, where $D=\epsilon_{0}E$ is the macroscopic polarisation (for linear response). Using Maxwell's equations, one may may compute $\Vec{\nabla}\times H=0$ and we already know $\Vec{\nabla}B=0$. From these two equations it becomes clear, that the magnetic field must be constant inside the perfect conductor. Therefore one speaks of a "frozen-in" magnetic field, since $B$ remains constant and depends on the history of the perfect conductor \cite{mcdonaldelectromagnetic}\cite{hofmann2015solid}. 

The situation is rather different in a SC. The SC's magnetisation does not depend on its previous history. If we apply a magnetic field only after cooling the SC below $T_{c}$, the SC behaves as expected. However, one if a SC is exposed to an electric field at room temperature and cooled beneath $T_{c}$, the SC still expels the external magnetic field: the Meissner-Ochsenfeld effect \cite{hofmann2015solid}.

\subsection{The theory behind \\  experiments}
Over the past century many theorists failed to explain the concept of superconductivity at a microscopic level. Great physicists like Einstein, Bohr, Kronig and Bloch failed to the extent that their predictions would fit the experiments \cite{Schmalian2010FailedTO}. However, eventually eye-opening achievements were made. Upon finding out about Meissner's and Ochsenfeld's discovery, the London brothers noticed the important relation between the electrons in the material and the applied magnetic field. Shortly after, Ginzburg and Landau introduced their famous phenomenological description of superconductivity. Bardeen-Cooper and Schrieffer (BCS) then succeeded in deriving a more general microscopic description of superconductivity in 1957\cite{BCS}.

\subsubsection{London equations}

Spurred by the discovery of the effect discovered by Meissner and Ochsenfeld, Fritz and Heinz London were pioneers in tackling superconductivity in a theoretical manner. Their starting point is classical electrodynamics. They started off with the free electron model being acted upon by the Lorentz force:

\begin{equation}
    E=\alpha \frac{dJ}{dt}
\label{London1}
\end{equation}

in which J, t, and E are current, time and electric field respectively. $\alpha=\frac{m}{ne^{2}}$, where m denotes mass, n is charge density and e is the electron charge. However there is a catch in, as F. and H. London called it, acceleration equation. As London and London admitted right from the beginning, equation \ref{London1} does not account for any friction of the charges at all \cite{london1935electromagnetic}.

Combining equation \ref{London1} with Faraday's law ($\vec{\nabla} \times E=-\frac{1}{c}\frac{dH}{dt}$) and since $\frac{1}{c}J=\vec{\nabla} \times H$ one arrives at the following equation\cite{london1935electromagnetic}:

\begin{equation}
    \vec{\nabla} \times \vec{\nabla} \times \alpha \frac{dH}{dt}= -\frac{1}{c}\frac{dH}{dt}
\end{equation}

Using $\vec{\nabla} H=0$ and integrating with respect to time yields:

\begin{equation}
    \alpha c^{2} \vec{\nabla}^{2}(H-H_{t=0})=H-H_{t=0}
\label{t0}
\end{equation}

They noted that the solutions of this inhomogenous differential equation behave ordinarily inside the SC, whereas one may compute that the solutions exponentially decrease as one moves further from the SC's surface into the external magnetic field $B$. They further argue that equation \ref{t0} gives a too general solution to the problem of the theoretical description of superconductivity. Therefore they replace $\vec{\nabla} \times H$ by $-\frac{1}{c}J$ resulting in the first London equation \cite{london1935electromagnetic}.

\begin{equation}
    \vec{\nabla} \times \alpha J=-\frac{1}{c}H
\label{l1}
\end{equation}

By solely working with equation \ref{l1}, they rid themselves of working with a very general differential equation and were able to simplify the abundant set of general solutions to natural behaviour, such as the Meissner effect \cite{london1935electromagnetic}. 

London and London further derived the second London equation in a similar fashion. 
Here they have used equations \ref{London1} and \ref{l1}. However, they found a component $\frac{\mu}{\alpha}$ which is proportional to the timelike component of the four-vector of $j_{i}$, the relativistic current density. Thus, they gave their differential equation a interpretation by the principle of covariance, thus an interpretation that observers in all reference frames may correlate the same physical quantities (introduced by Einstein \cite{Einstein:1916vd}). With this equation they postulated the replacement of Ohm's law ($J=\sigma E$, $\sigma$ being the conductivity) \cite{london1935electromagnetic}: 

\begin{equation}
    \alpha (\frac{dJ}{dt}+c^{2} \vec{\nabla} E)=E
\end{equation}

Finally, they also described a crucial physical property in SCs: the London-penetration depth ($\lambda$). Combining Ampere's law with equation \ref{l1}, one arrives at:

\begin{equation}
    \vec{\nabla}^{2} B=\frac{1}{c^{2}\alpha}B
\end{equation}

The solution to this differential equation is obvious. Expressing $(\frac{1}{c^{2}\alpha})^{-\frac{1}{2}}$ in SI units yields the London-Penetration depth, $\mu_{0}$ being vacuum permeability:

\begin{equation}
    \lambda_{L}=\sqrt{\frac{m}{\mu_{0}ne^{2}}}
\end{equation}

Even though they described some phenomena of this early stage of theoretical SC investigations quite well, they too were aware of the missing links of their formalism \cite{GORTER1934306}. They concluded that the electrons shield off the external magnetic field \cite{year100}.

\subsubsection{Ginzburg-Landau-Theory}
Soon thereafter further improvements were made by Vitaly Ginzburg and Lev Landau resulting in the Ginzburg-Landau theory in the 1950s. Their theory is based on the free energy of the superfluidity of electrons \cite{cyrot1973ginzburg}. Superfluidity is a quantum-mechanical state, in which particles lose all their friction \cite{schmitt2015introduction}. Using Landau's theory on second order phase transitions, they described superconductivity by a means of a complex order parameter $\psi$ of a second order phase transition related to the free energy of the system. This order parameter made it possible to determine to what extent the material is in the superconducting phase \cite{GLtheory}. It is important to note that any solutions to $psi$ are only non-zero solely in the superconducting phase (below a threshold temperature, denoted as the critical temperature $T_{c}$) \cite{ginzburg2009theory}. 

First they expanded the free energy $F$ of the system in terms of a complex order parameter:

\begin{equation}\label{free}
\begin{aligned}
F_{s}=F_{n}+\alpha \mid\psi\mid^{2}+\frac{\beta}{2}\mid\psi\mid^{4} \\
 +\frac{1}{2m}\mid\left(-i\hbar\vec{\nabla}-\frac{2eA}{c}\right)\psi\mid^{2}+\frac{1}{8\pi}\mid B\mid^{2}
 \end{aligned}
\end{equation}

$F_{n}$ denotes the normal free energy of the system (not being in the superconducting state), $m$ describes the mass of the electron. It becomes clearer later with BCS-Theory why the charge is $2e$ (Cooper-pairs) instead of $e$ (single electron). A is the vector potential and $B$ being the magnetic field. 

Using the Helmholtz free energy equation, one may now continue to determine the other variables, where $H_{c}$ denotes the magnitude of the energy needed to keep the magnetic field outside the material \cite{cyrot1973ginzburg}. 

\begin{equation}
    \frac{H_{c}^{2}}{8\pi}=F_{n}-F_{s}
\label{Helmholtz}
\end{equation}

In the absence of a magnetic field the equation simplifies to the first two terms (F. Assuming an equilibrium state $\frac{\partial F_{s}}{\partial\mid\psi\mid^{2}}=0$ being a minimum of free energy. Both $\alpha$ amd $\beta$ can now be readily computed using that $\mid\psi\mid^{2}_{T>T_{c}}=0$ \cite{ginzburg2009theory}. 

\begin{equation}
    \mid\psi\mid^{2}=\frac{-\alpha}{\beta}=\frac{T_{c}-T}{\beta}\frac{d\alpha}{dT}
\label{ab}
\end{equation}

It was known from experiments that equation \ref{ab} corresponded to the magnetic field $H^{2}$. Again, Ginzburg and Landau minimized the free energy $F_{s}$ by taking the derivative $\frac{d}{d\psi^{*}}$ and applying the boundary condition that neither the gradient of $\psi$ nor the direction of the vector potential acting on $\psi$ has parallel compontents at the boundary, which they call the natural boundary conditions. 

Again minimizing the free energy (with respect to $A$ and $\psi$, but this time using equation \ref{free}, one may now derive the two Ginzburg-Landau equations \cite{cyrot1973ginzburg} \cite{ginzburg2009theory}.

\begin{equation}
    \alpha \mid\psi\mid^{2}+\frac{\beta}{2}\mid\psi\mid^{4}+\frac{1}{2m}\left(-i\hbar\vec{\nabla}-\frac{2eA}{c}\right)\psi=0
\end{equation}
and
\begin{equation}
    \frac{c}{4\pi} \vec{\nabla} \times H=j= \frac{2e}{m}\left( \psi^{\ast}\left( -i\hbar \vec{\nabla} -\frac{2eA}{c}\right) \psi +h.c.\right)
\end{equation}

When using a hydrodynamic model to describe charge transport in a SC, one may now deduce more meaning from the wavelike function $\psi$. We can relate this variable to the number density: $n(r)=\mid \psi(r) \mid ^{2}$ with the current equation $j=2env$ \cite{cyrot1973ginzburg} or more precisely to the density matrix $\rho=\int n(r') dr'$ \cite{ginzburg2009theory}. 

That leads us right to the SCs' charge velocity:

\begin{equation}
    v(r)=\frac{1}{m}\left(\hbar\vec{\nabla} arg(\psi) -\frac{2eA}{c}\right)
\end{equation}

This approach further enhanced the understanding of the macroscopic behaviour of SCs such as the previously described type II SC. Using the Ginzburg-Landau theory, Abrikosov introduced magnetic field vertices that  entered the solid, thus allowing for mixed state SCs \cite{ABRIKOSOV1957199}. This mixed state entailed that the material would neither be a strictly superconducting nor an normal material.

\subsubsection{BCS-Theory}

While Ginzburg and Landau had focused heavily on the phenomenological description of superconductivity, Bardeen, Cooper and Schrieffer managed to describe the superconducting state on the microscopic level. They argued that, given low temperatures, electrons (spin 1/2) could pair up to so called Coooper-pairs and form quasi-bosons (energy particles). This electron pair wave function could then travel through matter freely, as low temperatures (low energy-states) annihilate electron-proton scattering effects \cite{BCS}.

Later called BCS-Theory, in 1957 Bardeen Cooper and Schrieffer sought to explain the second order phase transition at a critical temperature $T_{c}$, the experimental measurement of energy gaps for the electron excitations, the Meissner-Ochsenfeld effect, infinite conductivity as well as a dependence on the isotope. They built upon the Bloch model of conductivity and included phonon-electron interaction as well as Couloumb forces to their description of superconductivity as already proposed by Fröhlich \cite{frohlich1950theory}. They came to the conclusion that thermal electron-scattering needed to be addressed in order to theoretically describe these phenomena. In phonon-electron interaction they found electrons to be attracted to one another, if the difference between the electron states ($\Delta \epsilon$) were less than $\hbar \omega$: $\Delta \epsilon<\hbar \omega$. Bardeen Cooper and Schrieffer thus explained many already discovered physical concepts theoretically and finally received the Nobel Prize for their groundbreaking work in 1972\cite{BCS}\cite{pines19721972}. Gork'ov later derived the macroscopic Ginzburg-Landau theory from BSC-theory in 1959 \cite{gor1959microscopic}.

In terms of the description of unconventional -meaning Cooper-pairing outside of phonon exchange or non-BCS- SCs (USCs), many attempts and various solutions are at hand. However, theoretitians' computations for UCs (including TSCs) have largely been meager. The race for an adequate description of existing UCSs is therefore still ongoing \cite{stewart2017unconventional}.

\subsection{Superconducting Energy Gap}

One of the most central quantities in the discription of conventional SCs, especially for the experimental confirmation of superconductivity, is the superconducting energy gap $\Delta$ in the energy dispersion relation, which results from the Cooper-pairing mechanism, as the electrons condensate into the BCS ground state. They are thus no longer available to the fermionic energy spectrum\cite{hofmann2013einfuhrung}. To measure this temperature-dependent quantity, one oftentimes uses Angle Resolved Photoemission Spectroscopy (ARPES) or  Scanning Tunelling Microscopy (STM) (examples: \cite{eom2006persistent}\cite{rodrigo2004use}\cite{kitazawa1996superconducting}), which have become standard experimental condensed matter physics tools\cite{zhang2005experimental}\cite{meyer2021introduction}. The gap may also be measured directly, as demonstrated by Nicol et al. in 1960\cite{nicol1960direct}. One may also use heat capacity measurements to deduce $\Delta$\cite{hofmann2013einfuhrung}. 

Once $\Delta$ has been determined, one may use it to fit it to BCS-theory to yield characteristic superconducting phase-transition diagrams\cite{eom2006persistent}.

\subsection{Post-BCS Developments}

Along with first real-world applications of early superconductors, soon new kinds of material became of interest to the scientific community: partially filled 3d-, 4f-, and 5f-electronic shells and their chemical compounds in the 1960s. To describe it in short: Strongly Correlated Materials (SCMs). 

Initially, the most fascinating phenomenon in these novel materials was the metal-insulator phase transition within the group of transition metal oxides. Thereafter, heavy fermion systems (the effective mass being greater than the free electron's mass by some factor of hundreds to thousands) were discovered in the 1970s, which yielded interesting phenomena of phase transitions between magnetic moments and superconductivity as well as anomalous transport properties \cite{anisimov2010electronic}. One such heavy fermion SC is CeCu$_{2}$Si$_{2}$ with $T_{C}=0.79 K$ discovered by Steglich et. al in 1979\cite{steglich1979superconductivity}. In opposition to these strongly correlated systems, weak electron correlation may be found in classical band theory (semiconductors, simple metals), in which electron interaction is practically neglected\cite{anisimov2010electronic}. 

Theoretical models for these phenomena were necessary. An early approach was taken by the Englishman John Hubbard: His assumptions neglected electrons of the s- and p- shells, saying that they may be considered stationary around the nucleus, and only took into account the valence electrons of a given lattice atom. Furthermore, his theory only held for low temperatures (stationary lattice points) \cite{hubbard1964electron}. His method stems from the Slater-Koster tight-binding model of solids, which approximates the electrons's of a lattice point be confined within a small region around the atomic nucleus, thus minimally changing the wave function of the electron upon very limited interaction with other lattice points \cite{slater1954simplified}. 

An unique event was marked in 1986 by Bednorz and Müller by the discovery of a extraordinarily high $T_{C}$ of a cuprate \cite{bednorz1986possible}. They detected a $T_{C}$ of around 30 K. At first, the scientific community was sceptical of their results, however their experiments were confirmedly reproducable by other groups globally within several weeks. Some of them reported even higher $T_{C}$ between 40 and 80 K. Bednorz and Müller were awarded the Nobel Prize in Physics in 1987, just a year later \cite{buckel2008superconductivity}. Today, the highest measured $T_{C}$ of a SC was measured to be 133 K at atmospheric pressure\cite{schilling1993superconductivity} and 164 K at high pressure (31 GPa)\cite{gao1994superconductivity}.

New classes of SCs emerged in the late 2000s: The existance of iron-based SCs was verified in LaOFeAs by Hosono et. al with $T_{C}=26 K$ \cite{kamihara2008iron}. While the class of iron-based SCs has grown steadily \cite{PhysRevB.101.235163}\cite{PhysRevB.101.235163}\cite{PhysRevB.99.144514}\cite{LOOS201577}, around 10 years later, iron-free Pd-based SCs emerged in highly pressurized systems \cite{abdel2018high}. Almost simultaneously, a new conventional class of SC became of interest, namely hydride SCs. Since BCS-theory is well understood, new CSC can be found even today. High phonon frequencies, strong electron-phonon coupling as well as a high density of states is a promising recipe. Fullfilling these recquirements, looking at hydrides seems a sound path to take \cite{ginzburg1992once}, as shown in a sulfur hydride system with $T_{C}=203 K$ at high pressures \cite{drozdov2015conventional}. Many new room temperature, high pressure hydride SCs have been predicted in the last years \cite{sun2019route}\cite{peng2017hydrogen}\cite{song2021high}.

\subsection{Early applications}

While superconducting electromagnets do not purely retort the history of superconductivity itself, a small detour into that field shall be made, since today's commercial superconductors mainly arise in the form of superconducting electromagnets (such as magnetic resonance spectroscopy \cite{Nitz2016} and magnetic levitation trains \cite{maglev}). While research further continued in the ever-growing field of superconductivity, superconducting materials became readily available and more and more scientists became invested in superconducting electromagnets \cite{year100}. Early attempts to fabricate large electromagnets were however unsuccessful. Superconducting wires were wound up into coils. If the coils however scaled too large, the superconducting wire's ohmic resistance rose such that the temperature of the wire rose. Hence, the coil underperformed the scientists' expectations.  The heating process could be avoided by a loop of cooling the system with liquefied helium again and again called training \cite{year100}. Since the process was largely unpredictable \cite{stekly1965stable}, a new method was necessary. Stekly and Zar were able to stabilize such superconductor coils. In 1965, they found that if the initial current carrying capacity of the supercondutor exceeded its material-dependent maximum, it would "spill over" into the cooling substrate (helium). Furthermore, they were able to annihilate disturbances by introducing additional copper wires to the coil. If a disturbance appeared (magnetic flux, etc.) and part of the material returned from the superconducting to the normal state, these copper wires would conduct the current until the superconducting material recover to its superconducting state. This detour of the current is necessary as the normal conduction zone of the superconductor cannot recover on its own. This is due to the reduction in conductivity leading to a further disturbance in the form of an increase in heat output \cite{stekly1965stable}. In the early stages such superconducting coils were a main interest for early-stage particle accelerators \cite{smith1968pulsed}.

\section{Topological Super- \\ conductors (TSC)}

The field of topological superconductivity is particularly interesting, since it combines two fields of research: TMs and SCs. However, TSCs do not act according to BCS-theory. A regular SC has an adiabatic (smooth) transition between the BCS and the Bose-Einstein condensate of cooper pairs, which may be tuned by an interaction strength parameter. As discussed in TMs, we allow no adiabatic transition between two topologically different states. Hence, the classic theory of superconductivtity does not describe TSCs well enough and we need to look into more unconventional pairing mechanisms \cite{ando2015topological}.

If BCS theory does not hold, how does a TM become superconducting? The answer lies in the protected gapless excitations at the boundary of the material. Gapless meaning, if the momentum $k \rightarrow 0$ then also the energy $E\rightarrow 0$. It gets even more interesting if we consider that both time-reversal symmetry breaking TSCs (closely related to Quantum Hall Insulators) and time-reversal invariant TSCs (related to Topological Insulators) exist\cite{ando2015topological}. The mentioned gapless excitations however do not stem from a the typical electron/hole mechanisms but are electron/hole superpositions of Bogoliubov quasiparticles\cite{ando2015topological}, often referred to as Majorana fermions \cite{beenakker2014annihilation}. Conceptually, they may be described as being their own antiparticle in an excited state. In easier terms they are their own hole \cite{leijnse2012introduction}. Since Majorana fermions must have the same properties as their antiparticles, they may not possess electric charge or spin. Finally, odd-parity paring of the electron wave functions is essential\cite{li2019exploring}. Thus, three recquirements have to be fulfilled: First, we recquire odd-parity pairing symmetry with a superconducting
gap. Additionally, we need gapless surface states consisting of Majorana
fermions and finally we recquire MZMs in the superconducting vortex cores \cite{li2019exploring}.

As mentioned in the previous paragraph, some TSCs exhibit time-reversal symmetry breaking while other do not. We shall investigate how these two distinct TMs obtain their superconducting phase and name some real materials exhibiting such phases, to which one may refer. For the discussion of the TRI and TR SCs, we will refer to 2D systems, as a generalization can be made in a straightforward fashion, as in reference \cite{TRinvariantSC}.

\subsection{Classification of Topological Superconductors by Time \\ Reversal Symmetry}

\begin{figure*}[ht]
    \centering
    \includegraphics[width=1\textwidth]{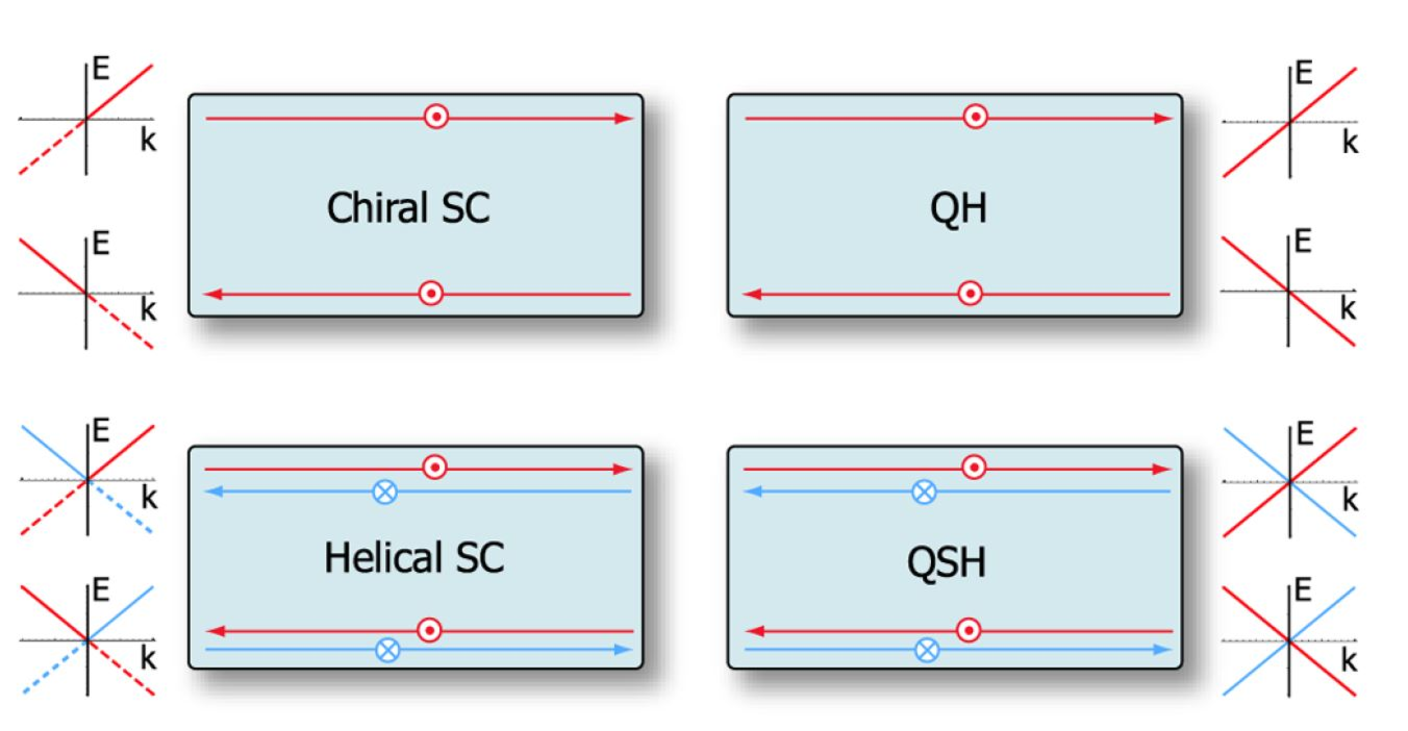}
    \caption{Schematic electron transport and dispersion relations of chiral and helical SCs opposing their non-superconducting Quantum Hall and Quantum Spin Hall relatives including their dispersion relations. Taken from \cite{TRinvariantSC}.}
    \label{Timereversal}
\end{figure*}

As already mentioned previously, time-reversal symmetry breaking TSCs exhibit Quantum-Hall like chiral edge states \cite{TRinvariantSC}. As shown in figure \ref{Timereversal}, time-reversal invariant TSCs show similar behaviour as in QSHE. The quasiparticle responsible for electron transport in topological insulators is the Dirac fermion, following the Dirac equation and being massless \cite{srednicki2007quantum} \cite{TRinvariantSC}. The quasiparticle underlying TRI SCs are helical chiral Majorana fermions, only having half the degrees of freedom as the Dirac fermions \cite{bjornson2016topological}. TRI SCs also channel up and down spins seperately opposing to TR symmetry breaking SCs, as can be seen in the figure.

Let us now consider the dispersion relations of the four materials at hand. The non-superconducting states (QHE and QSHE) show the spin channel behaviour well. Looking at the top right dispersion relation, it becomes clear that upon a TR operator acting on the state/system, the spins flip and the direction of momenta $k$ do as well, yielding the expected outcome: A linear dispersion relation will be reflected in the $k$-plane. Similarly, but with seperate spin channels (red and blue), the dispersion relation of the QSHE are also reflected, but with no overall change of the dispersion relation, since a reflection in the $k$-plane remains the same dispersion.

The superconducting dispersion relations (very left of figure \ref{Timereversal}) behave similarly, but only allow for positive energy values. For a detailed derivation from the Bogoliubov-de-Gennes Hamiltonian in the Nambu basis to the $E=-E$ and $k=-k$ equality in the dispersion relation of Majorana modes, the reader is adviced to study references \cite{bjornson2016topological} and \cite{bernevig2013topological}. In short, the reason for that phenomenon is that Majorana fermions are Kramer degenerate. If a TR operator $T$ acts on a state $\ket{\psi}$, the eigenenergy of the system remains the same \cite{Wigner1993UeberDO}. Even though, in general, their $\frac{1}{2}$-spin of electrons forbids the same quantum state, one should recall that we are occupied with Majorana fermions, making the negative energies redundant, as positive and negative solutions may be computed using the same k-values \cite{TRinvariantSC}. Majorana fermions do not depend on momenta greater than zero \cite{bernevig2013topological}.

\subsection{Classification by the Coupling Strength and Pairing \\  Mechanism}

Yet another distinction of TSC can be made by means of their coupling strengths and the Cooper paring mechanism. As already described in a previous chapter, the main goal to achieve superconductivity is to find some shielding mechanisms (often referred to as screening) that make originally repulsive electrons attract each other in some channel to form Cooper pairs. This is usually done via some angular momentum channel \cite{scaffidi2017weak}. Already in the 1960s, Kohn and Luttinger concluded that, in principle, superconductivity could endure even at minimal coupling strengths in a 3D electron gas with short interaction lengths\cite{kohn1965new}\cite{scaffidi2017weak}. This was later confirmed in numerous unconventional SC like cuprate SC and organic SC\cite{scaffidi2017weak}. 

A Strongly Coupled SC (SCS), examples being Pb and Hg\cite{scalapino1966strong}, on the other hand cannot be described in the same fashion. A Weakly Coupled SC (WCS), examples being Al and Zn \cite{scalapino1966strong}, may be described using Landau's quasiparticle approximation \cite{landau1959theory}, in which the strongly interacting electrons are replaced by weakly interacting quasiparticles near the Fermi surface \cite{neilson1996landau}. Within a SCS, Landau's quasiparticle approximation breaks down \cite{scalapino1966strong} and has to be replaced by a formulation of Eliashberg, in which Green's functions for electrons in a SC computed by Gork'ov come into play \cite{eliashberg1960interactions}. A classification via coupling shall not be confused by one via correlation strengths, which is of course also possible. A strongly correlated SC is for example Rh$_{17}$S$_{15}$ \cite{naren2008strongly}, while Mo$_{3}$Al$_{2}$C\cite{bauer2010unconventional} is considered a weakly correlated SCs. 

Despite such great success both in theory and experiment\cite{MAH}, when looking for TSC one shall not be too concerned with such conventional categorizations, since the mechanism at play are clearly not topological: The material transitions from a weakly-coupled BCS quantum state to a strongly coupled Bose Einstein condensate \cite{li2019exploring}. Therefore we are on the lookout for systems with unconventional electron-electron pairing mechanisms \cite{li2019exploring}. Especially, the combination of TRS-breaking and spin-triplet pairing in materials has promised more strong coupling TSC. To this day however, few materials of the kind exist\cite{li2019exploring}.

TSC may exhibt weak and strong electron-phonon and electron-electron coupling \cite{li2019exploring}\cite{scaffidi2017weak}. An example of a strong coupling TSC is the doped semimetal Mo$_{5}$Si$_{3}$ \cite{ruan2022strong}, wheras Sr$_{2}$RuO$_{4}$ is a good example of a weak coupling TSC \cite{scaffidi2017weak}. 

Not only the coupling strength, but also the coupling mechanism should be of interested when concerned with a TSC \cite{li2019exploring}. Conventional SCs' Cooper pairs usually pair via the s- or d-wave ($L=0,2$ respectively). The total spin is $S=0$, a singlet. TSC's pairing happens via triplet ($S=1$) pairing, which leads to p, f and h-wave superconductivity, because the overall wave function must be odd parity\cite{li2019exploring}.A prominent example of such an odd parity TSC is Cu$_{x}$Bi$_{2}$Se$_{3}$, which has received considerable attention in the previous decade\cite{PhysRevB.90.100509}\cite{sasaki2011topological}\cite{yonezawa2016bulk}. A more detailed classification of bulk SCs may be found in reference \cite{yonezawa2016bulk}.

\section{Conclusion and Outlook}

In conclusion, topological quantum materials have emerged as a fascinating and rapidly developing field of research at the intersection of condensed matter physics and materials science. These materials possess unique electronic properties and exhibit nontrivial topological characteristics that give rise to protected states and novel transport phenomena. The exploration of topological quantum materials has not only deepened our fundamental understanding of condensed matter physics but also opened up promising avenues for technological applications.The study of topological quantum materials has led to the discovery of various intriguing phenomena and the proposal of potential applications. Major advancements have been made in the understanding and manipulation of topological insulators, topological superconductors, and other topological phases. The discovery of Majorana zero modes, topologically protected surface states, and topological edge states has raised excitement for their potential use in topological quantum computing, fault-tolerant quantum gates, and robust information storage.Moreover, the investigation of topological quantum materials has led to the development of new techniques and experimental tools for characterizing and probing these materials. Scanning probe microscopy, angle-resolved photoemission spectroscopy (ARPES), and advanced transport measurements have played pivotal roles in the experimental study of topological quantum materials, allowing for the direct observation and manipulation of their unique electronic properties.Looking forward, the field of topological quantum materials holds great potential for further discoveries and technological breakthroughs. Ongoing research aims to uncover new topological phases, explore their properties in different material systems, and engineer novel functionalities. Additionally, efforts are being made to better understand the interplay of topological physics with other phenomena, such as strong correlations, magnetic order, and unconventional superconductivity.The development of topological quantum materials also benefits from the synergy between theory and experiment. Theoretical models and simulations, along with experimental validation, guide the search for new materials and aid in the understanding of their behavior. This interdisciplinary collaboration continues to drive the field forward and has the potential to lead to transformative advances in areas like quantum computing, spintronics, and energy-efficient electronics.In summary, topological quantum materials have revolutionized our understanding of condensed matter physics and hold great promise for technological applications. The field is still rapidly evolving, with ongoing discoveries, theoretical advancements, and experimental breakthroughs. The future of topological quantum materials is filled with excitement and the potential to reshape our technological landscape.

In a nutshell, we discussed the main causes of topological phenomena. The most fundamental mathematical tools to describe topological materials such as the Chern number, the principal concepts of time and parity symmetry as well as MZM were explained. With the attained knowledge, we were then able to understand the integer QHE as well as related effects. Such included the QAHE, QSHE but also the FQHE. We then deepened our knowledge on numerous topological materials. They often relate to the various Hall related effects, for example the QSHE in TI. Furthermore, relatively unexplored and novel materials such as CCs and MTIs became of substantial interest.

In the second part, we took a historical perspective on superconductivity, from Onnes discovery \cite{early} more than a hundred years ago to today's ongoing research in high pressure hydrides \cite{drozdov2015conventional}. We looked at the classification between type I and type II SCs. Besides experimental findings by Onnes \cite{early} and the Meissner-Ochsenfeld effect\cite{MOeffect}, ground-laying theoretical work as the London-equations\cite{london1935electromagnetic} and Ginzburg-Landau theory \cite{cyrot1973ginzburg} were discussed in brief, followed by the a short introduction to complete description of conventional superconductivity by Bardeen, Cooper, and Schrieffer\cite{BCS}. Finally, the fields of topology were brought in contact with superconductivity to form TSCs. TSCs were then classified by their TRS, coupling strength and pairing mechanism to give a better overview of that quickly emerging field.

With large data bases and advanced methods as DFT, it has become easier to find TSCs than previously possible by means of symmetry indicators. Such symmetry indicators may be found in the electronic structure of the material to detect non-trivial topological phases \cite{ono2019symmetry}. New kinds of TSCs are also popping up in various places. Examples include superconducting WSMs \cite{yan2017topological}, nodal TSCs driven by magnetic fields\cite{he2018magnetic} and artificial TSCs caused by Moiré patterns\cite{he2018magnetic}. Not only by their various foundations but also by their synthetisation method, advancements are to be expected: Hard tip point contact method, intercalation and doping, electric field gating are again only examples how  topological materials and more specifically TSCs may be realized \cite{li2019exploring}. Current topics of research include raising temperatures, ultrafast studies, exotic fermi arc behaviour\cite{slager2017dissolution}, CC with magnonic systems\cite{shindou2013topological}, optical response behaviour as well as the Jahn-Teller effect, which can help to stabilize MTIs \cite{ruegg2011topological}. 

While the main focus of applying TSCs is clearly quantum computing\cite{li2019exploring}, topological materials may become appealing to many different branches of engineering due to their low power dissipation. Examples include catalysis, energy conversion, data and energy storage as well as spintronics devices and meterology\cite{gilbert2021topological}\cite{pesin2012spintronics}\cite{poirier2009resistance}\cite{luo2022topological}\cite{li2021topological}. However, the number of applications those interesting materials shall have in the future, only time will tell.

\section{Acknowledgments}

MAH acknowledges support from the VR starting grant 2018-05339 and Wallenberg Foundation (Grant No. 2022.0079). 

\bibliography{Literature}

\end{multicols}

\end{document}